\documentclass[letterpaper,12pt]{article}
\usepackage{amsmath, amssymb}
\usepackage{epsfig}
\def\lsim{{\ \lower-1.2pt\vbox{\hbox{\rlap{$<$}\lower5pt\vbox{\hbox{$\sim$}}}}\ }}
\def\gsim{{\ \lower-1.2pt\vbox{\hbox{\rlap{$>$}\lower5pt\vbox{\hbox{$\sim$}}}}\ }}
\newcommand{\deltarpp}{{\mbox{{\small $\Delta_R^{++}$}}}}
\newcommand{\deltalpp}{{\mbox{{\small $\Delta_L^{++}$}}}}

\begin{document}

\begin{center}
 {\LARGE Prospects for the Search for a Doubly-Charged Higgs in the Left-Right Symmetric Model with ATLAS\\}

\bigskip
  G.~Azuelos$^{1,2}$, K.~Benslama$^3$ and J.~Ferland$^1$\\
\medskip
  $^1$ Universit\'e de Montr\'eal\\
  $^2$ TRIUMF, Vancouver\\
  $^3$ Nevis Labs, Columbia University

\bigskip

\LARGE{\today}

\end{center}

\bigskip
  We estimate the potential for observation at the LHC of a doubly
  charged Higgs boson, as predicted in Left-Right symmetric models. Single
  production by vector boson fusion, $W^+W^+ \to \Delta_{L,R}^{++}$
  and pair production by the Drell-Yan process $q \bar q \to
  \Delta_{L,R}^{++}\Delta_{L,R}^{--}$ are considered. Various decay
  channels are investigated: dileptons, including pairs of $\tau$'s,
  as well as $WW$.
\newpage

\section{Introduction}

One major puzzle of the Standard Model is the fact that weak
interaction couplings are strictly left-handed. In order to remedy
this apparent arbitrariness of nature, one must extend the gauge group
of the Standard Model to include a right-handed sector.  The simplest
realization is a Left-Right Symmetric Model (LRSM)~\cite{LRSM} based
on the group $SU(2)_L \otimes SU(2)_R \otimes U(1)_{B-L}$. Variants of
this model derive generally from Grand Unified Theories~\cite{GUT},
including superstring inspired models, based on extended groups which
contain the LRSM as a subgroup.  The breaking of $SU(2)_R \otimes
U(1)_{B-L} \to U(1)_Y$ occurs at a high energy scale due to a triplet
of complex Higgs fields
\footnote{Alternative minimal Left-Right symmetric models exist with
only doublets of scalar fields~\cite{ALR}. They do not lead to 
Majorana couplings of the right-handed neutrinos.}
, consisting of $\Delta_R^0$, $\Delta_R^+$ and $\Delta_R^{++}$, when
its neutral component acquires a non-vanishing vacuum expectation
value. The Higgs sector of the model therefore contains a doubly
charged Higgs boson, which could provide a clean signature at the LHC
since charge conservation prevents it from decaying to a pair of
quarks. This will be the subject of the present work.  For this
purpose, we refer to previous phenomenological
studies~\cite{Gunion89,Grifols89,Vega90,Huitu97}, expanding the
analyses by including the effects of backgrounds as well as detector
acceptance and resolution. It must be noted that very light,
${\cal{O}}(\sim 100)~ \mathrm{GeV}$, doubly-charged Higgs particles can be
expected in supersymmetric left-right models~\cite{Aulakh}.

The LRSM has a number of interesting features. The enlarged symmetry
group implies the existence of new heavy right-handed gauge bosons
$W_R$ and $Z_R$, while the fermion sector is enriched by representing
the right-handed fermions as doublets in $SU(2)_R$.  Yukawa couplings
of the triplet Higgs allow for Majorana mass terms of the right-handed
neutrinos, giving rise to the see-saw mechanism~\cite{seesaw} as a
natural explanation for the low, but non-vanishing mass of left-handed
neutrinos. These heavy neutrinos further provide a mechanism for
leptogenesis.

  Other signatures of the LRSM have been studied in ATLAS.  New heavy
$Z'$ gauge bosons can be observed in channels $Z' \to \ell^+ \ell^-$
up to masses of $~ 4.5$ TeV~\cite{Zprm,TDR} with an integrated luminosity
of 100 fb$^{-1}$.
In analyses specific to the LRSM, the search of $W_R \to \ell \nu_R$
 yields limits on the observability of the $W_R$ and right-handed
 Majorana neutrino~\cite{Wn}, as a function of the masses of these
 particles.  The channel $Z_R \to \nu_R \nu_R \to \ell q \bar q' ~
 \ell q'' \bar q'''$ should be observable~\cite{Zvv} up to masses
 $M_{W_R} \lsim$ 4 TeV and $m_{\nu_R} \lsim 1.2$ TeV with 300
 fb$^{-1}$.  The new gauge bosons can also be searched
 for~\cite{ZWWenjj,Arik} in the channels $Z_R \to WW\to e\nu jj$ and
 $W_R \to WZ$ if non-negligible mixing exists with the Standard Model
 (SM) gauge bosons. As a complement to these searches, observation of
 a doubly charged Higgs would clearly provide an important
 confirmation of the nature of the new physics.

  The Higgs sector~\cite{Gunion89} of the LRSM consists of (i) the
right-handed complex triplet $\Delta_R$ mentioned above, with weights
(0,1,2), meaning singlet in $SU(2)_L$, triplet in $SU(2)_R$ and $B-L =
2$, (ii) a left-handed triplet $\Delta_L$ (1,0,2) (if the Lagrangian
is to be symmetric under $L \leftrightarrow R$ transformation); and a
bidoublet $\phi$ (1/2,1/2,0).  The vacuum expectation values (vev) of
the neutral members of the scalar triplets, $v_L$ and $v_R$, break the
symmetry $SU(2)_L \times SU(2)_R \to U(1)_Y$ as well as the discrete
$L \leftrightarrow R$ symmetry.  The non-vanishing vev of the
bidoublet breaks the SM $SU(2)_L \times U(1)_Y$ symmetry. It is
characterized by two parameters $\kappa_1$ and $\kappa_2$, with
$\kappa = \sqrt{\kappa_1^2 + \kappa_2^2} =$ 246 GeV. To prevent
flavour changing neutral currents (FCNC), one must have $\kappa_2 \ll
\kappa_1$, implying minimal mixing between $W_L$ and $W_R$.  The mass
eigenstate of the singly charged Higgs is a mixed state of the charged
components of the bidoublet and of the triplet.  

In the analysis below, in order to have a manageable number of
parameters, we relate the mass of $W_R$ to $v_R$ by: $m_{W_R}^2 =
g_R^2v_R^2/2$, which is a valid approximation in the limit where $v_L
=0$ and $\kappa_1 \ll v_R$. All the above parameters have
bounds~\cite{Huitu97}: custodial symmetry constrains $v_L \lsim$ 9 GeV
and present Tevatron lower bounds on $M_{W_R}$ impose a limit $v_R >
1.4$ TeV, or $m_{W_R} > 650$ GeV, assuming equal gauge couplings $g_L
= g_R$. Direct limits from the Tevatron on the mass of the doubly
charged Higgs from di-leptonic decays have recently been reported
in~\cite{Tevatron}. Indirect limits on the mass and couplings of the
triplet Higgs bosons, obtained from various processes, are given
in~\cite{Swartz}.

The paper is organized as follows: The next section discusses the
phenomenology of the doubly charged Higgs at the LHC. Sect. 3
describes the methods of simulation of signals and background, as well
as detector effects. The following two sections give results on the
observability of the $\Delta_R^{++}$ and $\Delta_L^{++}$ respectively
in various decay channels. The last section summarizes and concludes.

\section{Phenomenology of the doubly-charged Higgs boson}\label{pheno}

\begin{figure}[!h]
\begin{center}
 \epsfig{figure=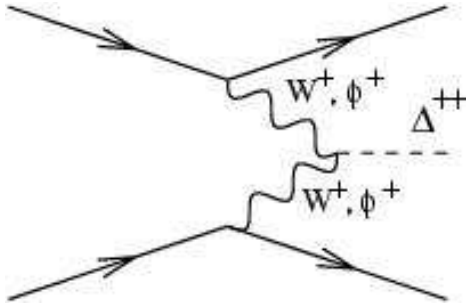,height=4.0cm}
 \bigskip
\caption{Feynman diagrams for single production of $\Delta^{++}$}
\label{Feynman}
\end{center}
\end{figure}

 Single production of a doubly charged Higgs at the LHC is possible
via vector boson fusion, or via the fusion of a singly-charged Higgs
with either a $W$ or another singly charged Higgs, as shown in
Fig.~\ref{Feynman}.  The amplitudes of the $W_LW_L$ and $W_RW_R$
vector boson fusion processes are proportional to $v_{L,R}$. The other
diagrams are suppressed by the Yukawa coupling of the Higgs to the
quarks and by a coupling $\phi\phi\Delta$ proportional to the small
parameter $v_L$.  For the case of $\Delta_R^{++}$ production, the
vector fusion process therefore dominates. For the production $W W \to
\Delta_L^{++}$, the suppression due to the small value of the $v_L$ is
somewhat compensated by the fact that the incoming quarks radiate a
lower mass vector gauge boson.

 Double production of the doubly charged Higgs is also possible via a
Drell-Yan process, with $\gamma$, $Z$ or $Z_R$ exchanged in the
$s$-channel, but at a high kinematic price since enough energy is
required to produce two heavy particles. In the case of
$\Delta_L^{++}$, double production may nevertheless be the only
possibility if $v_L$ is very small or vanishing.

The decay of a doubly charged Higgs can proceed by several channels.
Dilepton decay provides a clean signature, kinematically enhanced, but
the branching ratios depend on the unknown Yukawa couplings. Present
bounds~\cite{Huitu97,Swartz} on the diagonal couplings
$h_{ee,\mu\mu,\tau\tau}$ to charged leptons are consistent with values
{ $O$}(1) if the mass scale of the triplet is large. For the
$\Delta_L^{++}$, this may be the dominant decay mode if $v_L$ is very
small. One would then have a golden signature: $ q \bar q \to
\gamma^*/Z^*/Z'^* \to \Delta_L^{++} \Delta_L^{--} \to 4\ell$.  For
very low Yukawa couplings ($h_{\ell\ell} \lsim 10^{-8}$), 
the doubly charged Higgs boson could be
quasi-stable, leaving a characteristic dE/dx signature in the
detector, but this case is not considered here. The decay
$\Delta_{R,L}^{++} \to W_{R,L}^+ W_{R,L}^+ $ can also be significant.
However, it is kinematically suppressed in the case of
$\Delta_R^{++}$, and suppressed by the small coupling $v_L$ in the
case of $\Delta_L^{++}$.  Furthermore, reconstruction of $W_{L,R}$
pairs is difficult at the LHC since they don't produce a resonance and
since $W_R$ decays involving heavy Majorana neutrinos lead to complex
events.

 In the present work, we consider the production and decay modes
discussed above.  The results will be presented as limits in terms of
the couplings $v_L$ or $v_R$, taking fixed reference values for the
Yukawa couplings of the doubly charged Higgs bosons to the leptons. It
will then be a simple matter to re-interpret the results for different
values of these Yukawa couplings. We will assume a truly symmetric
Left-Right model, with equal gauge couplings $g_L = g_R = 0.64$. Since
the mass of the $W_R$ is essentially proportional to $v_R$, as
mentioned in the introduction, it will not be an independent
parameter.

We note that the existence of the Higgs triplet can also be detected
in the decay channel $\Delta^+ \to WZ$. This will not be studied here,
as the signal is very similar to narrow techicolor resonances which
have been analyzed elsewhere~\cite{techni}.


\section{Simulation of the signal and backgrounds}\label{simulation}

 The processes of single and double production of doubly charged Higgs
are implemented in the PYTHIA generator~\cite{PYTHIA}. Events were generated 
using the CTEQ5L parton distribution functions, taking account of
initial and final state interactions as well as hadronization. 
The following processes were studied here:

\begin{itemize}
 \item $W^+_{R,L} W^+_{R,L} \to \Delta^{++}_{R,L} \to e^+e^+/\mu^+\mu^+$
 \item $W^+_{R,L} W^+_{R,L} \to \Delta^{++}_{R,L} \to \tau^+\tau^+$ with one or both $\tau$'s decaying leptonically.
 \item $ q \bar q \to \gamma^*/Z_{R,L} \to \Delta^{++}_{R,L}\Delta^{--}_{R,L}$
\end{itemize}

The process $W^+_{R,L} W^+_{R,L} \to \Delta^{++}_{R,L} \to W^+_{R,L}
W^+_{R,L}$ has not been considered. The case of left-handed $W$'s was
previously studied~\cite{littleh} in the framework of the little Higgs
model and is discussed briefly in Sect.~\ref{WW}.

Detector effects and acceptance were simulated using
ATLFAST~\cite{ATLFAST}, a fast simulation program for the ATLAS
detector. Conditions of high luminosity (${\cal L} = 10^{34}$
cm$^{-2}$ s$^{-1}$) were assumed.  Jets were reconstructed from
calorimeter clusters using a cone algorithm, with a cone radius
$\Delta R = \sqrt{(\Delta\eta)^2+(\Delta\phi)^2} = 0.4$.  Charged
leptons (electrons and muons) were reconstructed within the acceptance
range in pseudorapidity $|\eta| < 2.5$. They were considered isolated
if they were separated by $\Delta R > 0.4$ from other clusters and if,
within a cone of radius 0.2, less than 10 GeV transverse energy was
deposited. An efficiency of reconstruction of 90\% was assumed for
these leptons. Reconstruction of $\tau$ jets assumed an efficiency of
40\% with a corresponding mistagging probability of $\sim$0.05\% for a
central $\tau$ jet having 100 GeV of transverse momentum. For b-jets,
the tagging efficiency was taken as 50\%, with a mistagging
probability of light quark jets of 1/230.

The principal backgrounds depend on the production and decay process
under consideration. PYTHIA was used to generate $t\bar t$ production,
which has a very large cross section of $~ 500$ pb. Other backgrounds
were simulated using the CompHep generator~\cite{CompHep}: (i) The
process $q q \to W^+W^+ q q$ was produced with cuts at generation
requiring transverse momentum of the outgoing quarks to be $> 15$ GeV
and an invariant mass of the $W^+W^+$ system greater than 200 GeV.
CTEQ5L parton density functions were used, with $Q^2 = m_W^2$. With an
assumed Higgs mass of 120 GeV, the contribution of longitudinal gauge
boson scattering is very small.  (ii) $q q \to W^+ Z q q$ was
generated under similar conditions. (iii) Finally the process $ p p
\to W t \bar t$ was also generated with CompHep. The cross sections of
these processes are given in Table~\ref{backgrounds}.

  A number of systematic uncertainties, some of which are difficult to
evaluate reliably before experimental data are available, will
apply. No k-factors have been used here, although
next-to-leading-order corrections can be substantial for these high
mass resonance states.  The luminosity measurement is expected to have
a precision of $\sim 5-10\%$. The efficiency of lepton reconstruction
in ATLAS, here taken to be 90\%, will have to be better understood.
The energy resolution, especially for high energy electrons, must be
evaluated with full detector simulation and optimized reconstruction
algorithms. Charge misidentification was not included here since it is
small ($\sim 5\%$ for leptons having a transverse momentum of 1 TeV).

\section{Search for \boldmath $\Delta_R^{++}$ \unboldmath}

  A different analysis strategy has been applied for each of the four modes of 
production and decay of the bileptons described in Sect.~\ref{pheno}. As mentioned
above, since the Yukawa couplings to leptons are not known, we will not account for
branching ratios into the various decay channels, but the reach on the parameter
$v_R$ could be recomputed trivially for any given values of these branching ratios.

  We will only consider signals for doubly positively charged Higgs bosons, as they
are about 1.6 times more abundant than the negatively charged ones, at the LHC 
(see Table~\ref{positive}). The same ratio of positively charged to negatively
charged leptons can be expected from the backgrounds, to the extent that
$qq WW$ dominates, and hence the improvement in the significances obtained below
can be estimated at 22\%. 

\begin{table}[h]
\begin{center}
\begin{tabular}{|c|c|}
\hline
       M($\Delta_{R}^{++}$)   & \% of ++ \\ 
\hline
\hline
   	300    & 64.5  \\ 
\hline
   	500    & 67.5  \\ 
\hline
   	800    & 71.5  \\ 
\hline
   	1000    & 75.3  \\ 
\hline
   	1500    & 79.1  \\ 
\hline
\end{tabular}
\caption{Percentage of positively charged $\Delta_R$ as a function of mass. A value
$m_{W_R} = 650$ GeV was assumed.}
\label{positive}
\end{center}
\end{table}

\subsection{Dilepton channel}
\label{sect:dilepton}

  The backgrounds relevant to the two-lepton channel are shown in
Table~\ref{backgrounds}.  A requirement of the presence of at least
two leptons ($e$ or $\mu$) has been applied, except for the $t\bar t$
sample, where only one lepton was required. Although the cross
sections are quite large, compared to those of the signal, given in
Table~\ref{tab:sBRll_R}, they can be effectively suppressed by
appropriate cuts, as will be seen below.

\begin{table}[h]
\begin{center}
\begin{tabular}{|c|c|c|}
\hline
    Background  &  ~Number of Events & ~$\sigma \times BR$ (fb)~  \\ \hline \hline
   	pp $\rightarrow$ W{\it t}$\bar{t}$                            &  200 000     &  23        \\ \hline
   	qq $\rightarrow$ $ W^+W^+~qq $                                &  100 000     &  37        \\ \hline
        qq $\rightarrow$ WZqq                                         &  27 000     & 28.6        \\\hline
        qq $\rightarrow$ {\it t}$\bar{t}$~~~ $P_t$ 10-200 GeV             &  8 000 000    &  90 800       \\ \hline
        qq $\rightarrow$ {\it t}$\bar{t}$~~~$P_t$ 200 GeV-$\infty$      &  2 000 000     & 14 100      \\ \hline
\end{tabular}
\caption{Background processes for the dilepton channel. The third column gives
$\sigma\times BR$, where $BR$ is the branching ratio to a final state with at
least two leptons, except for $t\bar t$ where only one lepton is required.
The number of events used for the analysis is also shown.}
\label{backgrounds}
\end{center}
\end{table}

\begin{table}[h]
\begin{center}
\begin{tabular}{|c|c|c|c|c|c|}
\hline
      M($W_{R}^{+}$)  &  \multicolumn{5}{c|}{~M($\Delta_{R}^{++}$)}\\
\cline{2-6}
               & 300   & 500   & 800   & 1000   & 1500 \\ \hline \hline
   	 650   &7.9   &4.6    &2.2    &1.4    &0.45   \\ \hline
   	 750   & 4.7  & 2.8   & 1.4   & 0.87  & 0.31  \\ \hline
   	 850   &2.9   &1.8    & 0.90  & 0.58  &0.21   \\ \hline
   	 950   &1.9   &1.2    &0.61   &0.40   &0.15   \\ \hline
   	 1000  &1.6   &0.98   &0.50   &0.33   &0.12   \\\hline
   	 1050  &1.3   &0.81   &0.42   &0.28   &0.11   \\\hline
   	 1500  &0.30  &0.20   &0.11   &0.074  &0.029  \\\hline
\end{tabular}
\caption{Cross section (fb) for the process 
$ p + p \to \Delta_R^{\pm\pm}$ as a function of
the mass of the $\Delta$ and of the $W_R$.}
\label{tab:sBRll_R}
\end{center}
\end{table}

 \subsubsection{$\Delta_R^{++} \to e^+e^+/\mu^+\mu^+$}
 \label{sec:rll}

\begin{figure}[h]
 \begin{center}
 \epsfig{figure=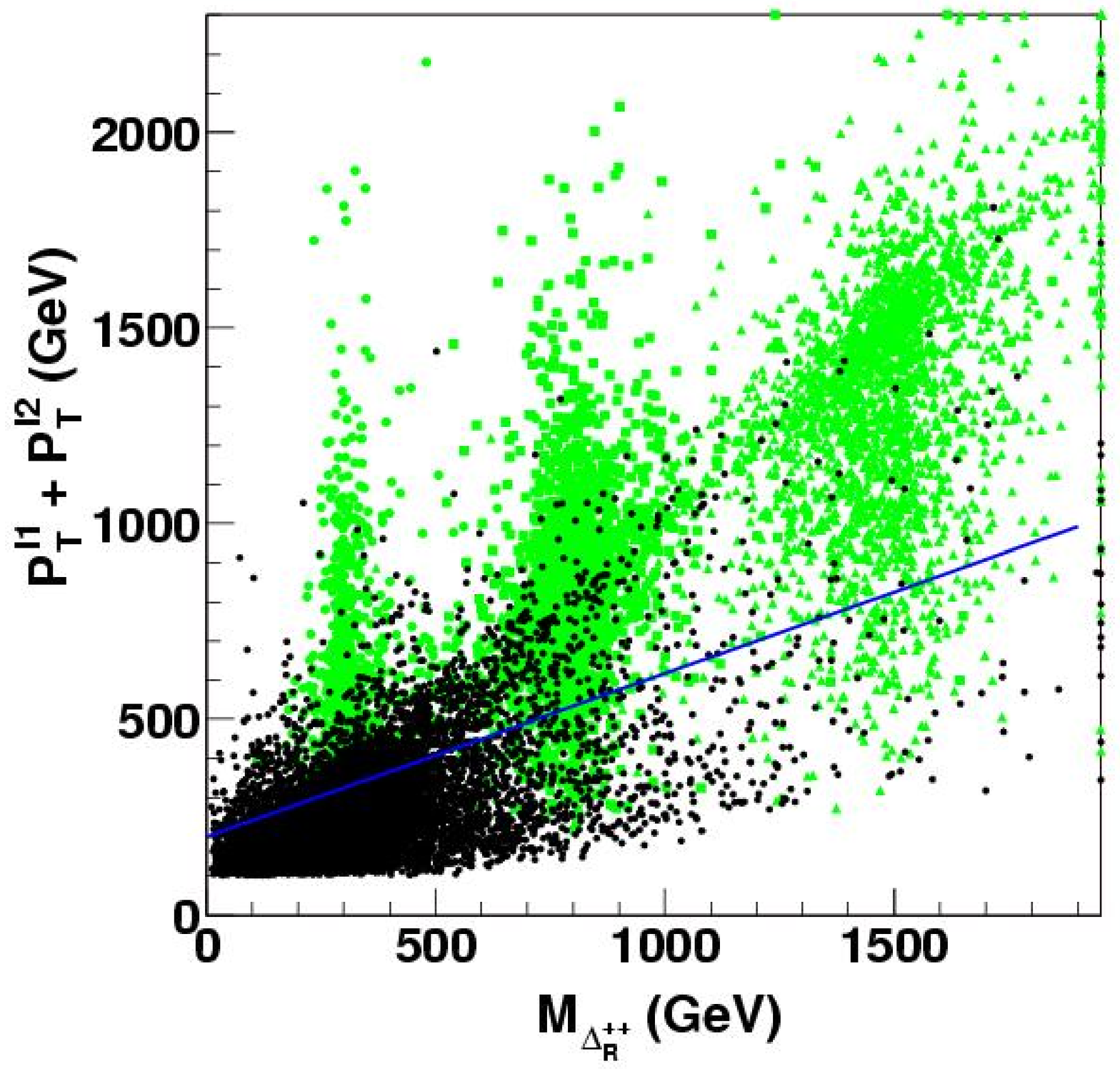,height=8.0cm}
 \caption{Distribution of the scalar sum of the two lepton transverse
   energies as a function of their invariant mass. In green (light shade) are
   shown the distributions for signals of $m_\deltarpp = 300, 800$ and 1500 GeV
          and $m_{W_R}$=650 GeV. 
   The sum of the backgrounds is in black. The straight line indicates where the
   mass-dependent cut is applied.}
 \label{fig:compar}
 \end{center}
\end{figure}

  The selection criteria for this channel were the following:
\begin{enumerate}
 \item Two isolated leptons ($\ell = e,\mu$) of positive charge were
 required. The ATLAS trigger requires that at least one of the leptons
 must have a transverse momentum $> 25$ GeV. The invariant mass of the
 two leptons, $M_{\ell\ell}$, was then computed.
 \item Since the leptons are expected to be energetic, an additional
requirement that both of them have a transverse momentum $p_T > 50$ GeV
was imposed. 
 \item The total transverse energy of the two leptons is correlated
with the mass of the \deltarpp\ from which they originate (see
Fig.~\ref{fig:compar}). A mass-dependent cut was therefore applied:
$\alpha (p_T^{\ell_1} + p_T^{\ell_2}) - M_{\ell\ell} = \beta$, where
the values $\alpha = 2.4$ and $\beta= 480$ GeV were chosen to suppress
background without significant loss of signal
 \item The two partons from which the vector bosons have been radiated
are expected to produce jets of high energy in the forward and
backward regions. A forward jet tag was therefore required:
 \begin{enumerate}
   \item After ordering all the jets with $p_T > 15$ GeV in decreasing
order of energy, the first one, $j_1$, was considered a candidate
forward-tagged jet if its energy was higher than 200 GeV.
   \item Looping over the remaining jets, the second candidate
forward-tagged jet, $j_2$, was identified if its angular separation
from $j_1$ satisfied $|\eta_{j_1}-\eta_{j_2}| > 2$ and if its energy
was greater than 100 GeV.
 \end{enumerate}
  The above requirements for ``forward jets'' are loose, but were 
found to be sufficient for the present purpose.
\item The missing transverse energy was required to be less than 100
GeV. This cut was not applied for cases $\deltarpp > 800$ GeV since it
was not needed to reduce the background.
\end{enumerate}

\begin{figure}[!h]
   \begin{center}
     \epsfig{figure=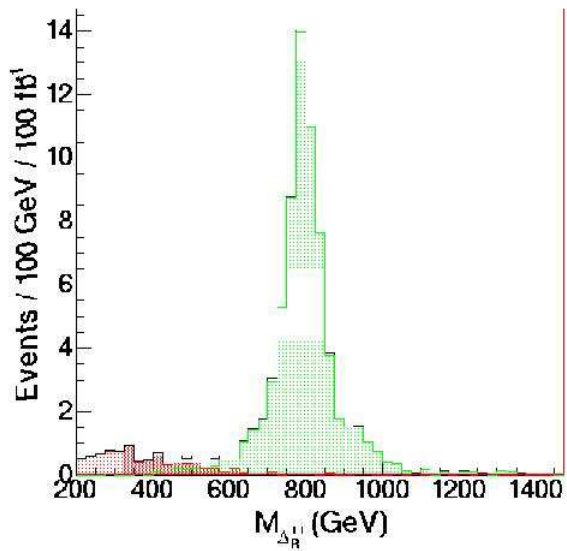,height=8.0cm}
     \bigskip
     \caption{Reconstructed invariant mass of the two leptons from the
process $W^+ W^+ \to \deltarpp \to \ell^+ \ell^+$. The signal (green)
is for a mass $m_\deltarpp = 800$ GeV with $m_{W_R}$ = 650 GeV and the
background is in red. The black histogram is the sum of both. The
distributions are for 100 fb$^{-1}$.}
     \label{fig:signal800}
   \end{center}
 \end{figure}

  Table~\ref{cuts-650-ll-R} shows the number of events expected from
the various backgrounds and for typical cases of signal where
$m_\deltarpp$ = 300 or 800 GeV and $m_{W_R^+} = 650$ GeV, after
successive application of the cuts. A window of $\pm 2\times$ the
width of the reconstructed mass of the \deltarpp\ was selected. The
numbers are normalized to a luminosity of 100
fb$^{-1}$. Fig.~\ref{fig:signal800} shows the distribution of
$m_\deltarpp$ and backgrounds for the case $m_\deltarpp$ = 800 GeV.

\begin{figure}[!h]
   \begin{center}
     \epsfig{figure=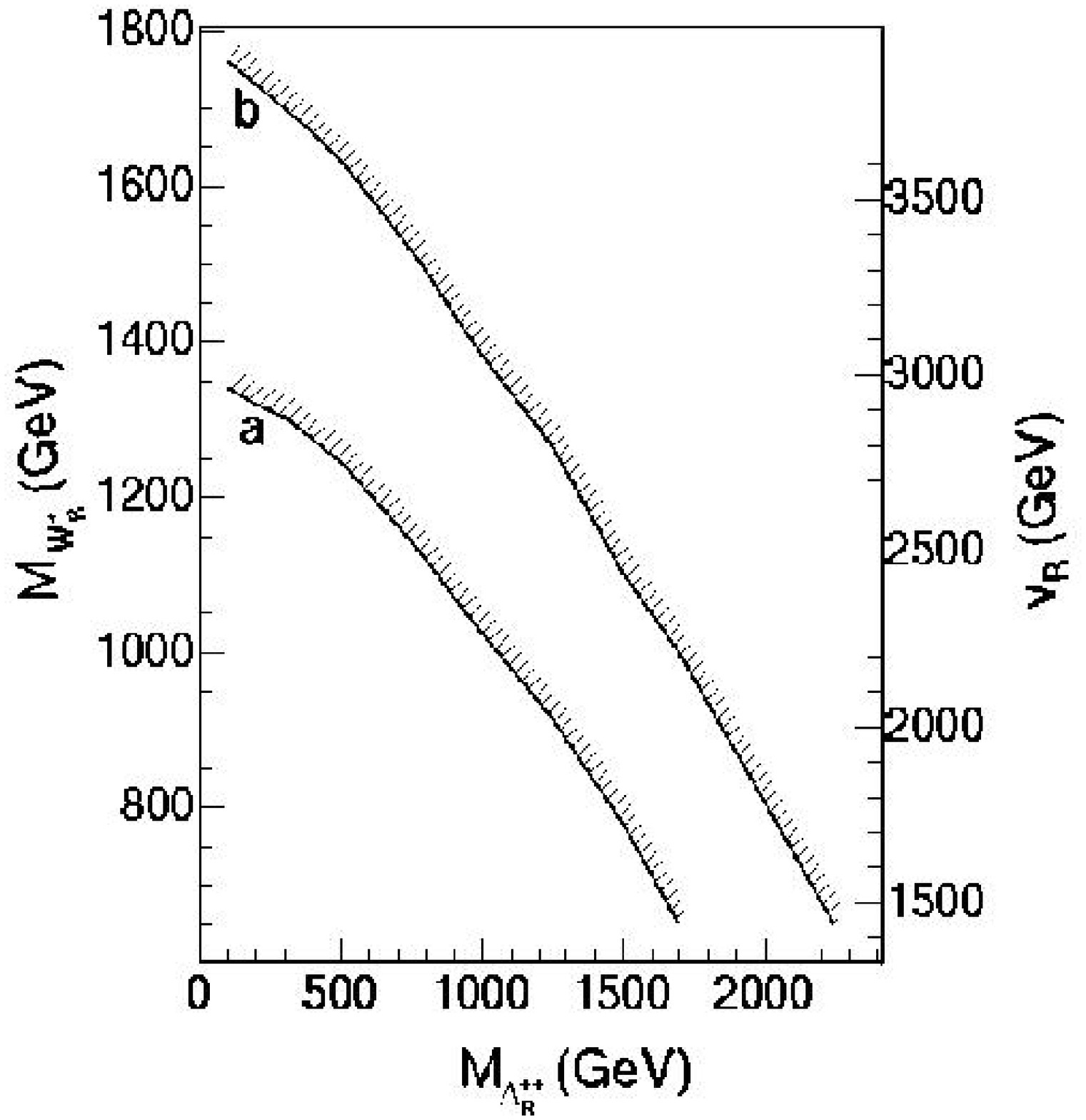,height=8.0cm}
     \bigskip
     \caption{Discovery reach for $\deltarpp \to \l^+l^+$ in the plane
$m_{W_R^+}$ versus $m_\deltarpp$ (or $v_R$) for integrated
luminosities of 100 fb$^{-1}$(a) and 300 fb$^{-1}$(b), and assuming 100\% BR
to dileptons.
}
     \label{fig:m_vs_m}
   \end{center}
 \end{figure}

 Since the background is negligible, discovery can be claimed if the
number of signal events is 10 or higher.  With this definition, the
contour of discovery, in the plane $m_{W_R^+}$ versus $m_\deltarpp$
(or $v_R$) has been estimated from a sample of test cases.  The
discovery reach at the LHC is shown in Fig.~\ref{fig:m_vs_m}, for
integrated luminosities of 100 fb$^{-1}$ and 300 fb$^{-1}$ and
assuming 100\% BR to lepton pairs.

\begin{center}
\begin{table}[htbp]
{\footnotesize
\begin{tabular}{|c|c|c|c|c|c|c|c|c|}
\hline
    	         &$\Delta^{++}$     &$\Delta^{++}$    & $W^+W^+~qq$  &$W~t\bar t$  & $WZqq$     &$t\bar t$      &total backg  \\ 
 	         &        300 GeV   &         800 GeV &              &             &            &               &             \\ \hline
Isolated leptons &   278 (327)      &  63 (95)        &     109/12   &    7.6/0.6  &    0/0.8   &      17/0    &     133/13  \\ \hline
Lepton $P_T$	 &   256 (301)      &  63 (94)        &     63/11    &    5.9/0.5  &     0/0.8  &     1.1/0    &     70/12   \\ \hline
 $2.4(P_T^{l_1}+P_T^{l_2})-M_{ll}>480$ &191(227) & 59(85) &    10/2.1&     1.3/0.3 &  0         &   0          &     12/2.4  \\ \hline
Fwd Jet tagging  & 156(186)         & 56(74)          &  6.0/1.3     &     0.1/0   &  0         &   0          &     6/1.3   \\ \hline
ptmiss           & 154(181)         & 56(68)          &    3.0/0.3   &     0/0     &  0         &   0          &     3.1/0.3 \\ \hline
\end{tabular}
}
\caption{Number of events of signal and backgrounds after successive
 application of cuts, for the case $\deltarpp \to \ell^+ ~\ell^+$, for
 $m_\deltarpp$ = 300 GeV and 800 GeV and $m_{W_R}$ = 650 GeV and for
 100 fb$^{-1}$. Mass windows $\pm 2 \sigma$ around the resonances have
 been chosen.  In parentheses is shown the number of events without
 the mass window cut.}
\label{cuts-650-ll-R}
\end{table}
\end{center}

 \subsubsection{\boldmath $\Delta_R^{++} \to \tau^+\tau^+$ \unboldmath }
\label{s:tautau}

 It is possible that the predominant decay of the \deltarpp\ will be
to the third generation leptons if the Yukawa couplings are
proportional to mass. For this important decay channel, the cleanest
signal of $\deltarpp \to \tau^+\tau^+$ will be the one where the two
$\tau's$ decay leptonically.  Although neutrinos are involved in the
process, the mass of the tau pair can be reconstructed, as was done
for example in $H/A \to \tau \tau$ studies~\cite{TDR}. One must
neglect the mass of the $\tau$ lepton and assume that the
neutrinos from the decay $\tau \to \ell \nu_\ell \nu_\tau$ are
collinear with the charged lepton. This is a good approximation to the
extent that \deltarpp\ is heavy and the $\tau's$ are highly
boosted. Defining $x_{\tau 1}$ and $x_{\tau 2}$ as the respective
fractions of tau energy carried by the charged lepton, conservation of
transverse momentum yields values for these variables and the
invariant $\tau \tau$ mass can then be calculated:

\begin{equation}
  m_{\tau \tau} = \frac {m_{\ell\ell}}{\sqrt{x_{\tau 1}x_{\tau 2}}}
\end{equation}

Since the energies involved here are large, it is not expected that
this method of $\tau \tau$ mass reconstruction will depend much on the
precision with which missing transverse energy can be measured, taken
here from fast simulation. 

The same method of reconstruction applies for 1-prong or 3-prong hadronic
decays of the $\tau$.

\bigskip
 {\underline {$\tau^+\tau^+ \to \ell^+\ell^+ p_T^{miss}$} }

 Assuming 100\% decay of \deltarpp\ to tau pairs, the cross section for
the process $ p + p \to \Delta_R^{\pm\pm} \to \tau \tau \to
\ell\nu\nu~\ell\nu\nu$ is the same as in Table~\ref{tab:sBRll_R}, but the
branching ratio BR($\tau \to \ell\nu\nu$) of 35\% per $\tau$ must be
taken into account. Besides the backgrounds of Table~\ref{backgrounds},
we have also taken into account $Zjj$, for which details on the generation
can be found in~\cite{mazini}.

 The following selection criteria were applied to extract the signal
 from the backgounds:
\begin{enumerate}
 \item two isolated leptons of the same charge were required in the event,
with $p_T > 25$ GeV. The event trigger is thus satisfied.
 \item the calculated variables $x_{\tau 1}$ and $x_{\tau 2}$ were
required to lie within the physical range $0 \le x_{\tau i} \le
1$. Events for which sources of missing energy other than the
neutrinos from the two tau decays are present, or for which the
collinear approximation made above is not valid, will be lost.
 \item in order to reduce the $t\bar t$ background, the event was rejected
if a $b$-tagged jet was present.
 \item forward jet tagging was required as in Sect.~\ref{sec:rll}
 \item Missing transverse momentum was required to be at least 150 GeV
\end{enumerate}

\begin{table}[htbp]
\begin{center}
{\footnotesize
\begin{tabular}{|c|c|c|c|c|c|c|c|c|}
\hline
    	         &$\Delta^{++}$     &$\Delta^{++}$    & $W^+W^+~qq$  &$W~t\bar t$  &$WZqq$      &$t\bar t$      &total backg  \\ 
 	         &        300 GeV   &         800 GeV &              &             &            &               &             \\ \hline
Isolated leptons &   44 (49)        &  20 (23)        &     153/80   &   13/5.1    &   12/0.7   &   486/137    &    707/234  \\ \hline
$0 < x_{\tau 1},~x_{\tau 2} <1$& 44 (46)    &  20 (21)&     84/60    &   6.4/3.3   &   8.0/0.7  &   360/101    &    480/171  \\ \hline
no b-jet         & 42 (44)          & 18(21)          &    84/59     &   0.2/0.2   &   7.2/0.7  &   62/16      &    175/83   \\ \hline
Fwd Jet tagging  & 36 (34)          & 16 (18)         &  21/15       &    0/0      &   2.3/0    &   20/7.4     &    45/23    \\ \hline
$E_T^{miss} > 150$ GeV& 23 (25)     & 13 (15)         &  5.2/7.2     &     0/0     &   0.8/0    &   2.5/2.1    &    8.6/9.3  \\ \hline
\end{tabular}
}
\caption{Number of events of signal and backgrounds after successive
 application of cuts, for the case $\deltarpp \to \tau^+\tau^+ \to
 \ell\nu ~\ell\nu$, for $m_\deltarpp$ = 300 GeV and 800 GeV and
 $m_{W_R}$ = 650 GeV, for 100 fb$^{-1}$. Mass windows $\pm 2 \sigma$
 around the resonances have been chosen.  In parentheses is shown the
 number of events without the mass window cut.}
\label{cutsR-650-tt-ll}
\end{center}
\end{table}

\begin{figure}[!h]
   \begin{center}
     \epsfig{figure=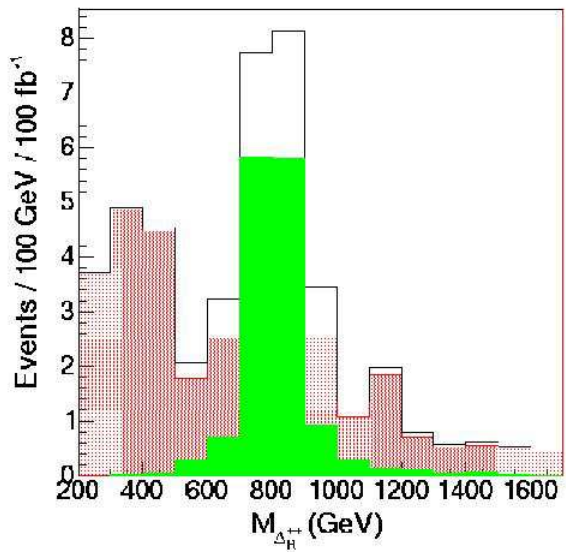,height=8.0cm}
     \bigskip
     \caption{Reconstructed mass of the \deltarpp\ from the
decay channel $\deltarpp \to \tau^+\tau^+ \to \ell^+ \ell^+ +P_T^{miss}$.
In red (light shade) and green (dark shade) are shown the background 
and the signal, for an integrated luminosity of 100 fb$^{-1}$. The
solid black histogram is the sum.}
     \label{fig:mtau}
   \end{center}
 \end{figure}

\begin{figure}[!h]
   \begin{center}
     \epsfig{figure=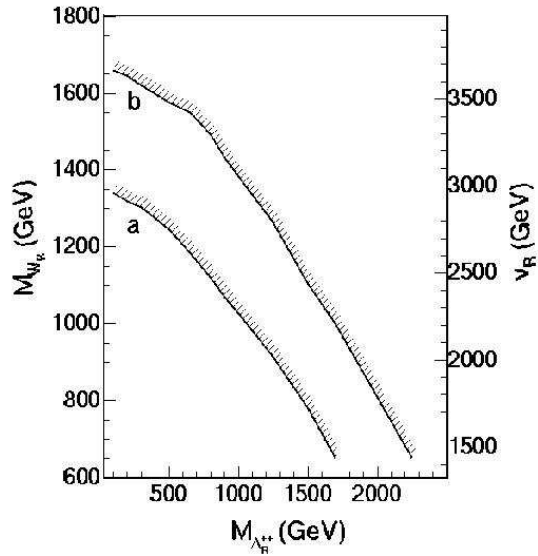,height=8.0cm}
     \bigskip
     \caption{Discovery reach in the plane $m_{W_R}$ vs $m_\deltarpp$
(or $v_R$) in the decay channel $\deltarpp \to \tau^+\tau^+ \to \ell^+
\ell^+ +P_T^{miss}$ for integrated luminosities of 100 fb$^{-1}$(a) and
300 fb$^{-1}$(b).}
     \label{fig:R_tau_limit}
   \end{center}
 \end{figure}

  Table~\ref{cutsR-650-tt-ll} shows the number of events expected from
the various backgrounds and for the cases of signals where
$m_\deltarpp$ = 300 and 800 GeV and $m_{W_R^+} = 650$ GeV, after
successive application of the cuts.  A window of $\pm 2\times$ the
width of the reconstructed mass of the \deltarpp\ was selected.
Fig.~\ref{fig:mtau} shows the distribution of $m_\deltarpp$ and
backgrounds for $m_\deltarpp = 800$ GeV.  A significance $S/\sqrt{B}$
of only about 4.3 is obtained in this case for an integrated
luminosity of 100 fb$^{-1}$. With simultaneous search for the
$\Delta_R^{--}$, and with 300 fb$^{-1}$, the significance should reach
9.1. Contours of discovery in this channel, defined as a 5$\sigma$
significance with at least 10 events, is shown in
Fig.~\ref{fig:R_tau_limit}. We note that they are weaker than for
$\deltarpp \to \ell^+\ell^+$ of Fig.~\ref{fig:m_vs_m}, and do not
cover a large region of mass which is unconstrained ($m_{W_R} > 650$
GeV). However, it may help to confirm discovery, or may be more
applicable if the coupling to $\tau$'s dominates.

\bigskip
 {\underline {$\tau^+\tau^+ \to \ell^+ h p_T^{miss}$} }\\
  Analysis of this channel shows that the background from $W$ + jets will
completely dominate the signal. Therefore, it will not be discussed here, but
details can be found in~\cite{these}.

 \subsection{Pair production \boldmath $\Delta_R^{++}\Delta_R^{--} \to 4 \ell$ \unboldmath}
\label{sec:pair}
\begin{figure}[!h]
   \begin{center}
     \epsfig{figure=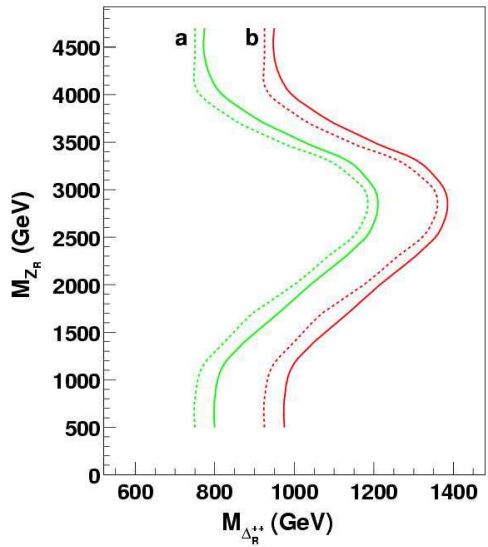,height=8.0cm}
     \bigskip
     \caption{Contours of discovery for 100 fb$^{-1}$(a) and for 300
fb$^{-1}$(b) in the plane $m_{Z'}$ vs $m_\deltarpp$. The dashed curves
are for the case where all four leptons are observed, and the full
curves are when only three leptons are detected.}
     \label{fig:pair_H_R}
   \end{center}
 \end{figure}

Pair production of $\Delta_R^{++}\Delta_R^{--}$ is suppressed by the
expected high mass of the \deltarpp\ but can nevertheless serve to
confirm discovery in some region of mass.  The diagrams with $s$-channel
$Z$ and $Z'$ exchange have been added to the $\gamma$ exchange diagram
in the implementation of the Drell-Yan process in the PYTHIA
generator, taking the coupling of $Z,~Z'$ to fermions and to
\deltalpp\ from references~\cite{Cuypers,Grifols89}. In principle,
the branching ratio depends on the assumed mass of $\Delta_L^{++}$,
as well as that of $\Delta_R^{++}$, but since the $Z'$ has a large partial
width to fermions, such that $BR(Z' \to \Delta^{++}\Delta^{--})$ is 
of the order of 1\%, the contribution of these decay channels to
the total width of the $Z'$ was neglected.  For the case of
leptonic decays of the doubly-charged Higgs bosons, 
the process constitutes a golden channel and the
background will be negligible. Fig.~\ref{fig:pair_H_R} shows the
contours of discovery, defined as observation of 10 events, if all
four leptons are detected or if any 3 of the leptons are
observed. Being an $s$-channel process not involving the $W_R$, this
channel is not sensitive to the mass of this heavy gauge boson.

 \subsection{Other channels}
  Other possible channels have not been considered here.  Single
production followed by the decay {$\Delta_R^{++} \to W^+_RW^+_R$} is
possible if $m_\deltarpp$ is sufficiently large, but given the lower
bound on $m_{W_R}$, this channel is strongly suppressed
kinematically. It would be also difficult to reconstruct since the
decays $W_R \to \ell N$ would not lead to a mass resonance and would
require a knowledge of the mass of $N$, a heavy right-handed or
Majorana neutrino.  Pair production $\Delta_R^{++}\Delta_R^{-}$ and
other such combinations can arise from $s$-channel $W$ exchange and can
complete the study of Sect.~\ref{sec:pair}

\section{Search for \boldmath $\Delta_L^{++}$ \unboldmath }

  \subsection{Dilepton channel}

  As for the case of the \deltarpp\, the dilepton channel provides a
clean signature. Although the Yukawa coupling of \deltalpp\ to leptons
remains a parameter of the theory, this channel can, in fact, be
dominant since the alternative decay to gauge bosons is possibly
negligible, being proportional to the very small value of the vev
$v_L$. In the limit where $v_L =0$, it will be the only open channel,
but production of \deltalpp\ will only occur in pairs, through
$s$-channel $\gamma/Z/Z'$ exchange. As before, we will assume below
100\% branching ratio to leptons, but results can be reinterpreted in
a straightforward way for different values of this branching ratio.

 The production cross section of \deltalpp\ is proportional to
 $v_L^2$. Table~\ref{tab:sBR_ll} gives the value as a function of
 mass, for $v_L$ = 9 GeV.

\begin{table}[h]
\begin{center}
\begin{tabular}{|c|c|c|c|c|c|c|}
\hline
      \multicolumn{6}{|c|}{~M($\Delta_{L}^{\pm\pm}$)}\\
\hline
  300 & 400 & 500 & 600 & 700 & 800 \\ \hline \hline
  9.847&6.155 &4.057 &2.822 &2.024 &1.494  \\ \hline
\end{tabular}
\caption{Cross section (fb) for the process 
$ p + p \to \Delta_L^{\pm\pm}$ as a function of
the mass of the $\Delta$ for $v_L = 9$ GeV.}
\label{tab:sBR_ll}
\end{center}
\end{table}

     \subsubsection{\boldmath $\Delta_L^{++} \to e^+e^+/\mu^+\mu^+$ \unboldmath}

\begin{figure}[hbtn]
\begin{center}
\mbox{\epsfig{file=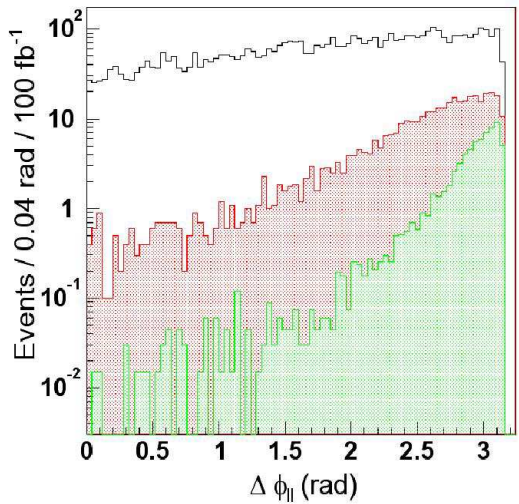,height=7.5cm}}
\caption{$\Delta \Phi_{ll}$ for a \deltalpp\ of mass 300 GeV in red
(dark shade) and for a mass of 800 GeV in green (light shade), as well
as the standard model background.  }
\label{f:dphill}
\end{center}
\end{figure}

\begin{figure}[!h]
\begin{center}
 \epsfig{figure=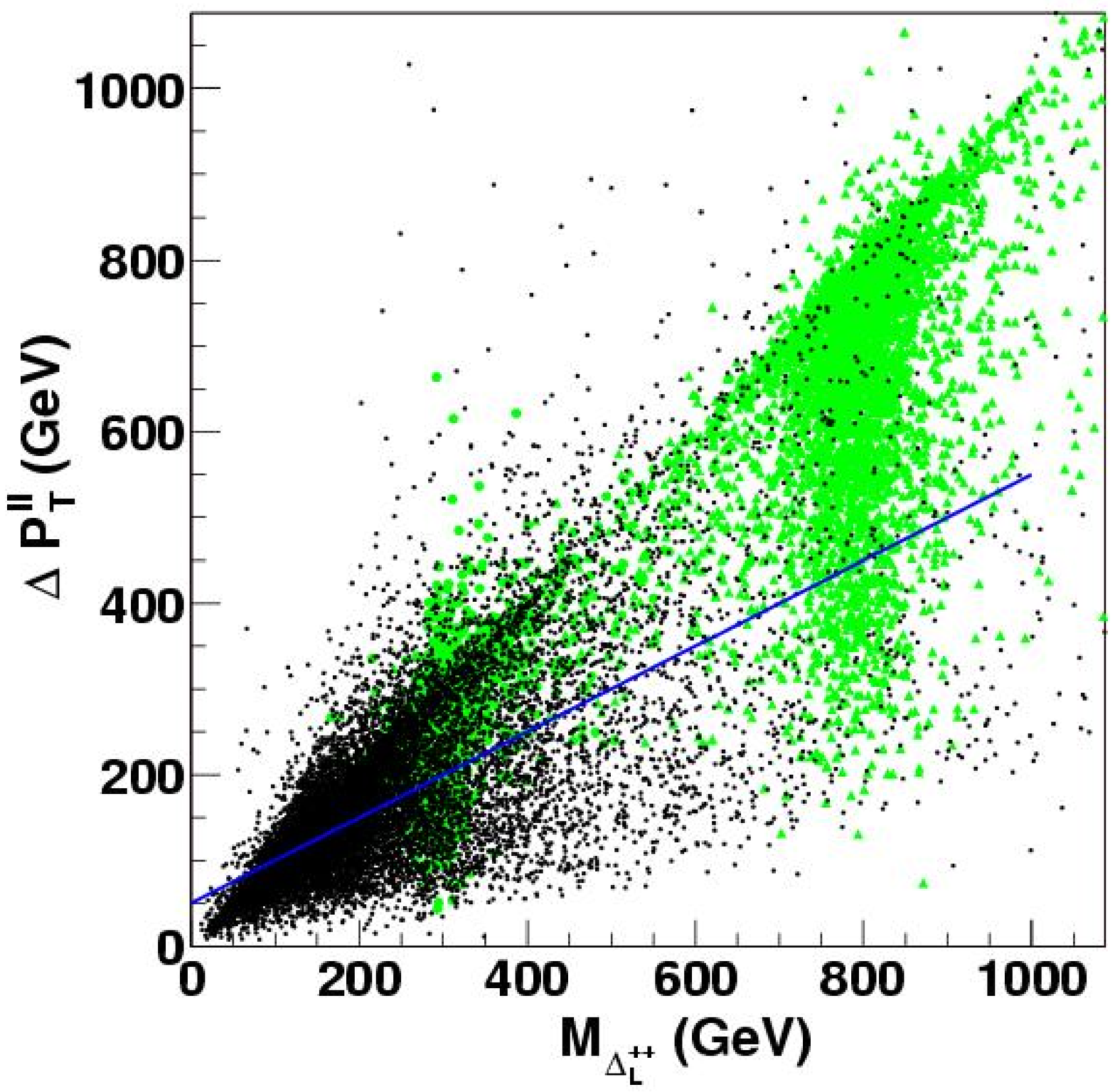,height=8.0cm}
 \bigskip
 \caption{Distribution of the scalar sum of the two lepton transverse
   energies as a function of their invariant mass in the case
   $\Delta_L^{++} \to \ell\ell $. In green (light shade) are shown the
   distributions for signals of $m_\deltalpp = 300$ and 800 GeV.  The
   sum of the backgrounds is in black. The straight line indicates
   where the mass-dependent cut is applied.}
 \label{fig:dptll}
\end{center}
\end{figure}

The following selection criteria were applied:
\begin{enumerate}
  \item Two isolated leptons ($e$ or $\mu$) of positive sign must be
reconstructed, with $p_T > 25$ GeV
  \item Since the leptons result from the decay of a heavy, slow
  particle, they will be collinear. The azimuthal separation between
  the two leptons is required to satisfy the cut $\Delta \Phi_{ll} >
  2.5$ when the reconstructed mass is 300 GeV or more. For lower
  masses, we choose $\Delta \Phi_{ll} > 1$ (see Fig.~\ref{f:dphill}).
\item The vectorial difference between the transverse momenta of the
  two leptons ($\Delta_{P_{T}^{ll}}$)is tuned for each assumed mass of
  the reconstructed \deltalpp. We require: $\Delta_{P_{T}^{ll}}$ $>$
  ($\frac{M_{ll}}{2}$+50) for masses 200 GeV or more and
  $\Delta_{P_{T}^{ll}} > 70$ GeV otherwise (see Fig.~\ref{fig:dptll}).
    \item Forward jet tagging is applied as in Sect.~\ref{sec:rll}.
  \item The missing energy is required to be less than 40 GeV since,
  in the SM, a same-sign lepton pair will always be associated with
  missing energy from accompanying neutrinos. In addition, we reject
  events where one jet is tagged as a b-jet.
\end{enumerate}

\begin{figure}[!h]
\begin{center}
 \epsfig{figure=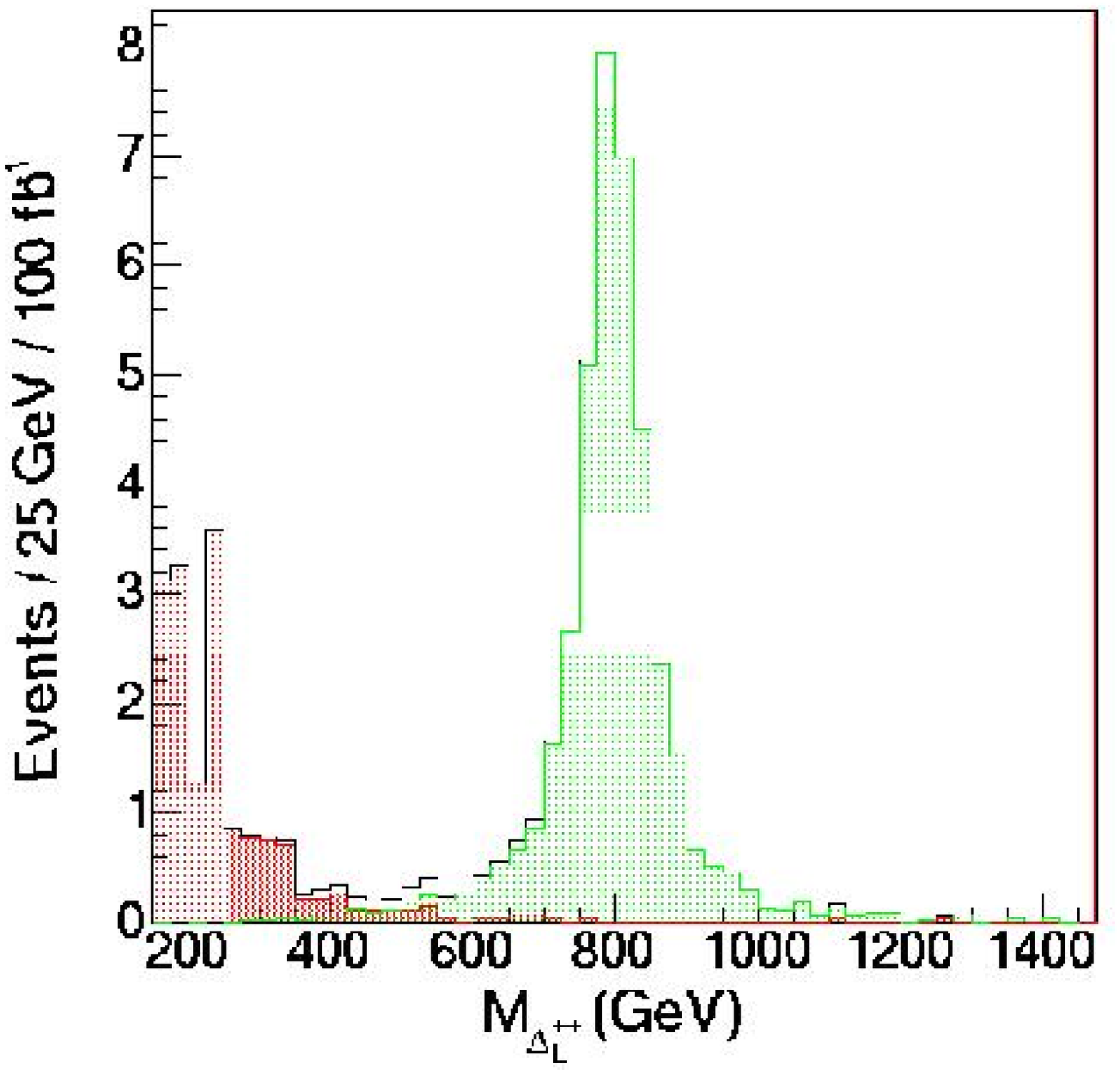,height=8.0cm}
 \bigskip
 \caption{Reconstructed invariant mass of the two leptons from the
process $W^+ W^+ \to \deltalpp \to \ell^+ \ell^+$. The signal (green)
is for a mass $m_\deltarpp = 800$ GeV and the
background is in red. The black histogram is the sum of both. The
distributions are for 100 fb$^{-1}$ and $v_L$ = 9.}
 \label{signal_hl}
\end{center}
\end{figure}

\begin{figure}[!h]
\begin{center}
 \epsfig{figure=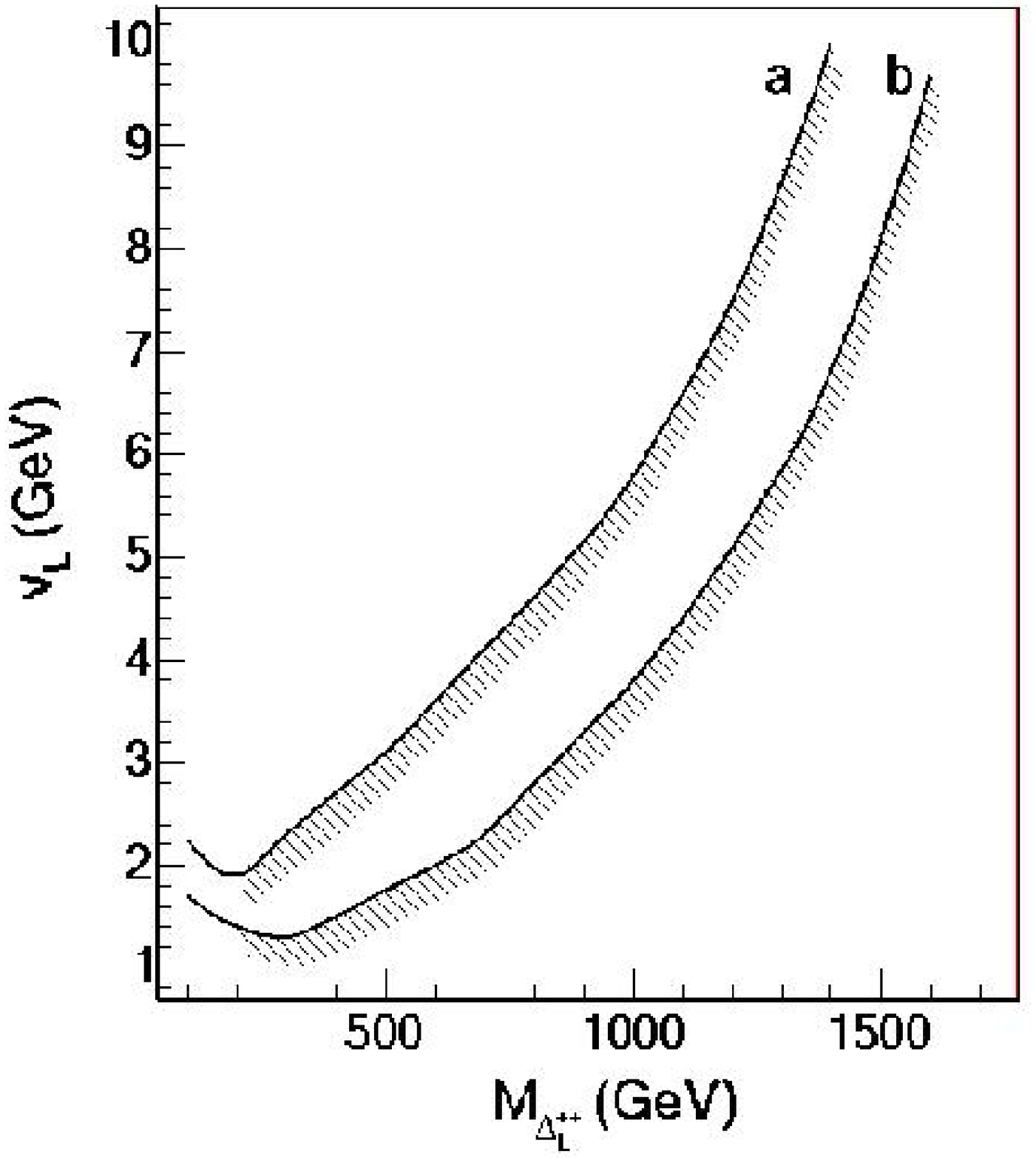,height=8.0cm}
 \bigskip
 \caption{Discovery reach for $\deltalpp \to \ell^+\ell^+$ in the
plane $v_L$ versus $m_\deltalpp$ for integrated luminosities of 100
fb$^{-1}$(a) and 300 fb$^{-1}$(b) and assuming 100\% BR to dileptons. }
 \label{limit_HL_lep}
\end{center}
\end{figure}

 Table~\ref{t:nev_L} gives the number of
 expected signal and background events for the cases $m_\deltalpp =
 300$ GeV and $m_\deltalpp = 800$ GeV respectively. A
 mass window of $\pm 2\times$ the width of the resonance was selected. 
 An example of a signal over SM background is given in Fig.~\ref{signal_hl}.
The discovery reach in the plane $v_L$ vs $m_\deltalpp$ is shown in
Fig.~\ref{limit_HL_lep}. 

\begin{table}[htbp]
\begin{center}
\begin{tabular}{|c|c|c|c|}
\hline
    	                                          &$\Delta^{++}$     &$\Delta^{++}$    &total backg  \\ 
 	                                          &        300 GeV   &         800 GeV &             \\ \hline
Isolated leptons                                  &   330 (384)      &     59 (69)     &  133/13     \\ \hline
$|\Delta\phi_{\ell\ell} > 2.5|$                   &   253 (289)      &     56 (65)     &  75/8.3     \\ \hline
$\Delta_{P_{T}^{ll}}$ $>$ ($\frac{M_{ll}}{2}$+50) &  220 (260)       &     50 (59)     &  37/2.5     \\ \hline
Fwd Jet tagging                                   & 156(185)         &     40 (47)     &  17/1.4     \\ \hline
ptmiss                                            & 152(180)         &     34 (40)     &  3.0/0.1    \\ \hline
\end{tabular}
\caption{Number of events of signal and total background after
 successive application of cuts, for the case $\Delta^{++} \to \ell^+
 ~\ell^+$, for $m_\deltalpp$ = 300 GeV and 800 GeV and $m_{W_R}$ = 650
 GeV, for 100 fb$^{-1}$. Mass windows $\pm 2 \sigma$ around the
 resonances have been chosen.  In parentheses is shown the number of
 events without the mass window cut.}
\label{t:nev_L}
\end{center}
\end{table}

     \subsubsection{\boldmath $\Delta_L^{++} \to \tau^+\tau^+$ \unboldmath}

 As mentioned above, the Yukawa coupling of \deltalpp\ to $\tau$
leptons may dominate, in which case it will be essential to
reconstruct the decay $\deltalpp \to \tau^+\tau^+$.  As for the case
of \deltalpp\ presented above, besides the backgrounds of
Table~\ref{backgrounds}, we have also taken into account $Zjj$.

\medskip
 {\underline {$\tau^+\tau^+ \to \ell^+\ell^+ p_T^{miss}$} }\\
The following cuts have been applied to select the di-lepton final states:

\begin{enumerate}

\item two same sign leptons with $P_{T} > $ 25 GeV.
\item Tau reconstruction:
      $0 <x_{\tau1} < 1$ and $0 <x_{\tau2} < 1$, where
      $x_{\tau1,\tau2}$ are defined as in Sect.~\ref{s:tautau}
\item b-jet veto: a b-jet tagging efficiency of 0.6 is applied
\item forward jet tagging, as in Sect.~\ref{sec:rll}
\item in order to reduce the $t \bar{t}$ and $qqWW$ backgrounds
an additional cut was applied on the invariant mass of
the two leptons: $M_{l1l2} > 30$ GeV
\end{enumerate}

\begin{figure}[h]
   \begin{center}
     \epsfig{figure=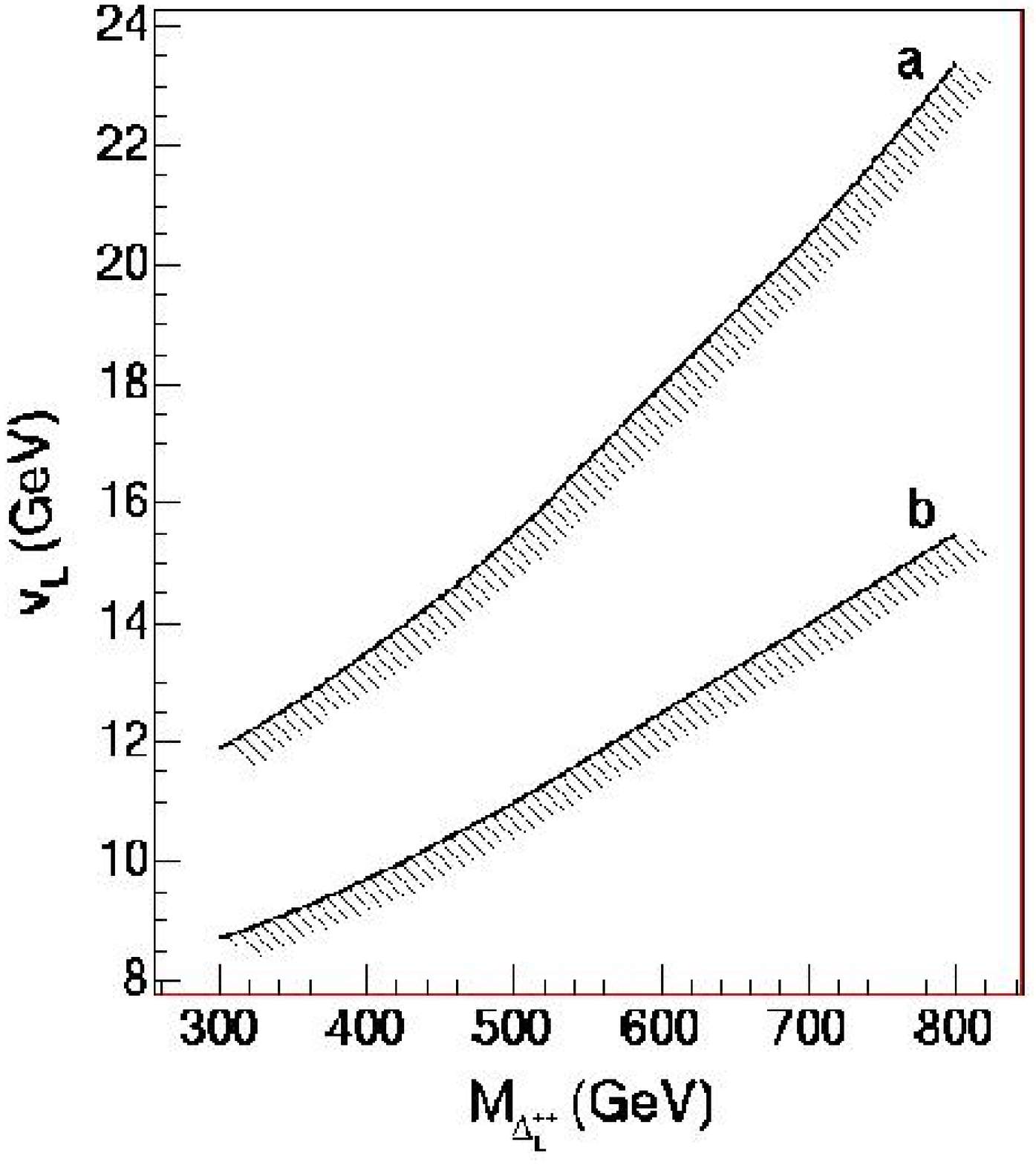,height=8.0cm}
     \caption{Discovery reach, in the plane $v_L$ versus $m_\deltalpp$
for integrated luminosities of 100 fb$^{-1}$(a) and 300 fb$^{-1}$(b) and
assuming 100\% BR to dileptons. 
}
     \label{limite1}
   \end{center}
 \end{figure}

Results of the analysis for a \deltalpp of mass of 300 GeV and 800
GeV are summarized in Table~\ref{t:ttl}.  A mass 
window of $\pm 2\times$ the width of the reconstructed resonance has
been selected. The parameter region where discovery is possible with 100 fb$^{-1}$ is
shown in Fig.~\ref{limite1}.

\begin{table}[htbp]
\begin{center}
\begin{tabular}{|c|c|c|c|c|c|c|c|c|}
\hline
    	                                         &$\Delta^{++}$     &$\Delta^{++}$    &total backg  \\ 
 	                                         &        300 GeV   &         800 GeV &             \\ \hline
Isolated leptons                                 &   42 (54)        &  10.1(13.4)     &    707/299  \\ \hline
$0 < x_1,~x_2 <1$                                &   40 (47)        &  9.6 (12.0)     &    480/222  \\ \hline
no b jet                                         &   38 (46)        &  9.1 (11.5)     &    158/100  \\ \hline
Fwd Jet tagging                                  &   18 (22)        &  4.3 (5.8)      &     33/23   \\ \hline
$M_{ll} > 30$ GeV                                &   15 (16.3)      &  3.8 (4.8)      &      23/16  \\ \hline
\end{tabular}
\caption{Number of events of signal and backgrounds after successive
 application of cuts, for the case $\deltalpp \to \tau^+\tau^+ \to
 \ell\nu ~\ell\nu$, for $m_\deltalpp$ = 300 GeV and 800 GeV and $v_L$
 = 9, for 100 fb$^{-1}$. Mass windows $\pm 2 \sigma$ around the
 resonances have been chosen.  In parentheses is shown the number of
 events without the mass window cut.}
\label{t:ttl}
\end{center}
\end{table}

\medskip
{\underline {$\tau^+\tau^+ \to \ell^+ h p_T^{miss}$}}\\
 As for the case of \deltarpp\ discussed above, this channel is dominated
by $W$ + jets background and will not been shown here. Details can be
found in~\cite{these}.

  \subsection{\boldmath $\Delta_L^{++} \to W^+W^+$ \unboldmath}
\label{WW}
  The decay of \deltalpp\ to a pair of $W$ bosons has been previously
  analyzed in the context of the little Higgs model~\cite{littleh} and
  will not be repeated here. It was found that the background from the
  $qqW^+W^+$ production process (with transverse $W$'s) was significant
  and that a 1 TeV resonance could be discovered at the LHC in this
  channel, with an integrated luminosity of 300 fb$^{-1}$, only if
  $v_L >$ 29 GeV. This value of $v_L$ is higher than would be
  reasonably expected from the constraints on the $\rho$ parameter, as
  mentioned in the introduction.

  \subsection{Pair production $\Delta_L^{++}\Delta_L^{--}$}
\begin{figure}[!h]
   \begin{center}
     \epsfig{figure=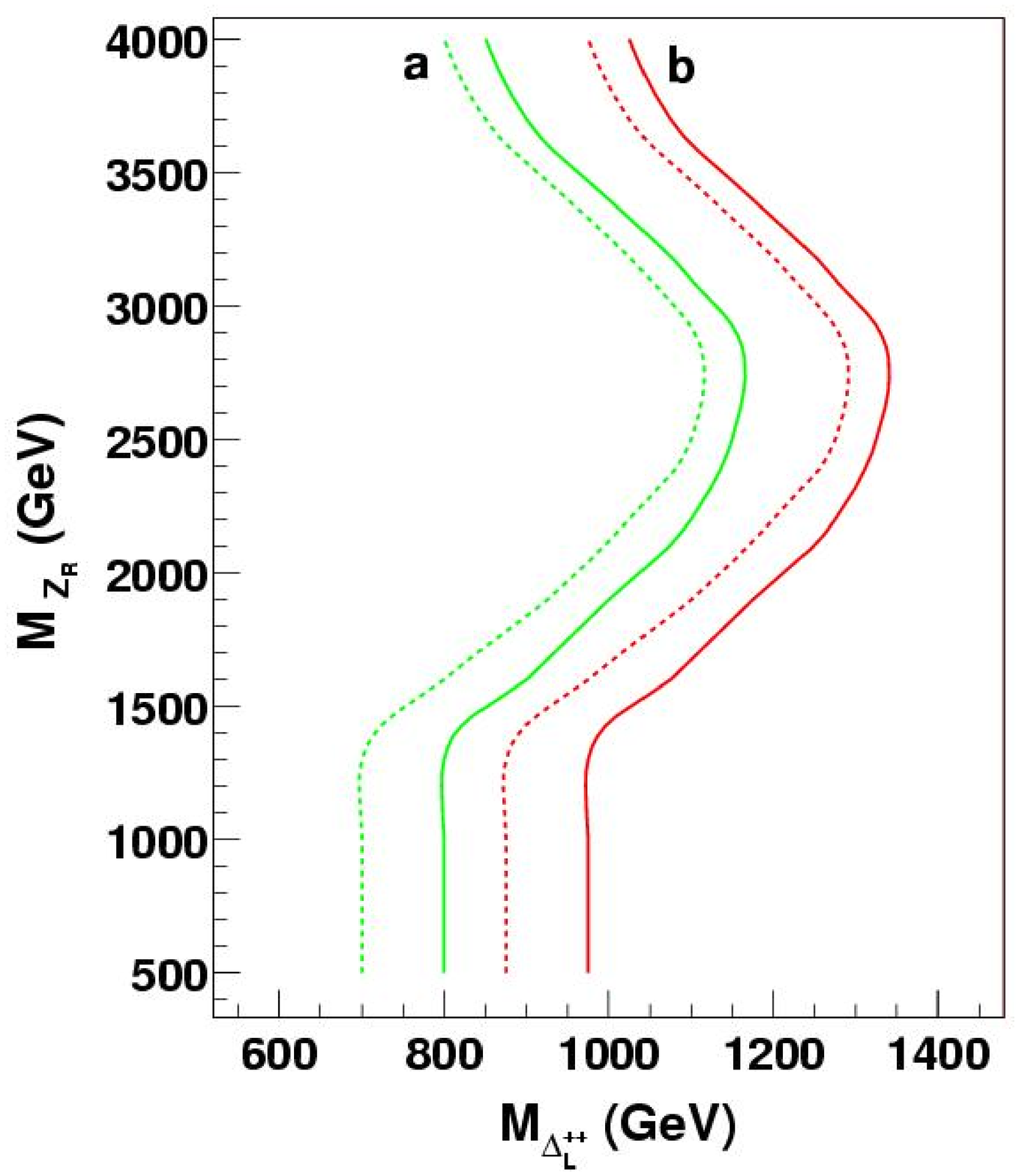,height=8.0cm}
     \bigskip
     \caption{Contours of discovery in the plane $m_{Z'}$ vs
$m_\deltalpp$ for 100 fb$^{-1}$(a) and 300 fb$^{-1}$(b). The dashed curves
are for the case where all four leptons are observed, and the full
curves are when only three leptons are detected.}
     \label{fig:pair}
   \end{center}
 \end{figure}

  As for the case of the right-handed sector, pair production of $\Delta_L$ is a possible 
discovery channel. The diagram with $s$-channel $Z'$ exchange has been added to the
implementation of this Drell-Yan process in the PYTHIA generator, taking the coupling
of $Z'$ to fermions and to \deltalpp\ from references~\cite{Grifols89,Cuypers}. Assuming
leptonic decays, the background will be negligible. Fig.~\ref{fig:pair} shows the
contours of discovery, defined as observation of 10 events, if all four leptons are
detected or if any 3 of the leptons are observed.

\section {Summary and Conclusion}

  Left-Right symmetric models predict the existence of doubly-charged
  Higgs bosons which should yield a striking signature at the LHC. The
  principal possible production and decay modes have been
  investigated and the reach of ATLAS for discovery is summarized in
  Figs.~\ref{fig:m_vs_m}, \ref{fig:R_tau_limit}, \ref{fig:pair_H_R}, \ref{limit_HL_lep}
  and~\ref{limite1} in terms of the parameters of the model. These
  plots are not independent. As the couplings to fermions are not
  known, the different channels have been considered separately,
  assuming 100\% branching ratio in each case.

  It is found that the LHC will be able
  to probe a large region of unexplored parameter space in the triplet
  Higgs sector. This analysis complements previous ATLAS studies
  searching for signals of the Left-Right symmetric model.

\section{Acknowledgments}
   This work has been performed within the ATLAS collaboration. We
have made use of physics analysis and simulation tools which are
the result of collaboration-wide efforts. 
We would like to thank Pierre-Hugues Beauchemin and Nathaniel Lubin
   for their help. K.B. acknowledges support from the National Science
   Foundation USA,  and G.A. from NSERC Canada.

\end{document}